\documentstyle[12pt, aaspp4]{article}
\lefthead{WOOSLEY, EASTMAN, \& SCHMIDT}
\righthead{}
\def\gtaprx {\lower .1ex\hbox{\rlap{\raise .6ex\hbox{\hskip .3ex
	{\ifmmode{\scriptscriptstyle >}\else
		{$\scriptscriptstyle >$}\fi}}}
	\kern -.4ex{\ifmmode{\scriptscriptstyle \sim}\else
		{$\scriptscriptstyle\sim$}\fi}}}
\def\ltaprx {\lower .1ex\hbox{\rlap{\raise .6ex\hbox{\hskip .3ex
	{\ifmmode{\scriptscriptstyle <}\else
		{$\scriptscriptstyle <$}\fi}}}
	\kern -.4ex{\ifmmode{\scriptscriptstyle \sim}\else
		{$\scriptscriptstyle\sim$}\fi}}}
\def \sun {$_{\scriptscriptstyle \odot}$}
\begin{document}
\begin{center}
Submitted to {\em The Astrophysical Journal}
\end{center}
\title{Gamma-Ray Bursts and Type Ic Supernova: SN 1998bw \\ }

\author{S. E. Woosley$^{a,b}$, Ronald G.  Eastman$^{a,b}$, and Brian
P. Schmidt$^c$  \\ }
\vskip 0.2 in
\affil{$^a$Lick Observatory, University of California Observatories,
Santa Cruz, CA 95064 \\ }
\affil{$^b$Lawrence Livermore National Laboratory, Livermore CA,
94550 \\ }
\affil{$^c$Mt. Stromlo and Siding Spring Observatories, Australian National
University}
\authoremail{woosley@ucolick.org}

\begin{abstract} 
Recently a Type Ic supernova, SN 1998bw, was discovered coincident
with a gamma-ray burst, GRB 980425. The supernova had unusual radio,
optical, and spectroscopic properties. Among other things, it was
especially bright for a Type Ic and rose quickly to maximum. When
modeled in the usual way as a spherically symmetric explosion,
this requires a large mass of $^{56}$Ni, 0.45 - 0.60 M\sun, a quite
massive star, and a very large explosion energy. We explore here
models based upon helium stars in the range 9 - 14 M\sun \ and
carbon-oxygen stars 6 - 11 M\sun \ which experience unusually
energetic explosions (kinetic energy $0.5 - 2.8 \times 10^{52}$
erg). Bolometric light curves and multi-band photometry are calculated
and compared favorably with observations. No spectroscopic data are
available at this time, but both LTE and non-LTE spectra are
calculated for the model that agrees best with the light curve, a
carbon-oxygen core of 6 M\sun \ exploded with a kinetic energy of $2.2
\times 10^{52}$ erg.  We also examine potential mechanisms for
producing the observed gamma-ray burst (GRB) - shock break-out and
relativistic shock deceleration in circumstellar material. For
spherically symmetric models, both fail to produce a GRB of even the
low luminosity inferred for GRB 980425.  The high explosion energies
required to understand the supernova are in contrast to what is
expected for such massive stars and may indicate that a new sort of
explosion has been identified, possibly the consequence of a collapsar
(Woosley 1993, 1996) whose main sequence mass was $\sim 35$ M\sun \ (helium
core mass 14 M\sun). Indeed a more likely explanation for what was
seen is a highly asymmetric explosion in which the GRB was produced by
a relativistic jet, perhaps viewed obliquely, and only a fraction of
the total stellar mass was ejected, the remainder accreting into a
black hole. The ejected mass (but not the $^{56}$Ni mass), explosion
energy, and velocities may then be smaller. Other associations between
luminous Type Ic supernovae and GRB's may exist and should be sought,
but most Type Ib and Type Ic supernovae do not make GRB's.

\end{abstract}


\section{Introduction}

Gamma-ray bursts (GRB's) have been a challenge to theorists and a source of
fascination for all for over 30 years and many models have been suggested to
explain them (Nemiroff 1993). Lately major progress has occurred in
understanding GRB's because of accurate localizations provided
by the Beppo-Sax mission. These locations allow rapid follow-up
observations with optical, x-ray, and radio telescopes that have
yielded exciting information about GRB counterparts. Two bursts have
been found to lie in galaxies having red-shifts of 0.83 and 3.42 and
are inferred to have had enormous energies, $\sim$10$^{52}$ erg and
$\sim 3 \times 10^{53}$ erg for GRB 970508 and GRB 971214 respectively. It
is currently believed that most gamma-ray bursts occur at such great
distances that their mean energy is at least 10$^{51}$ erg {\sl in
gamma-rays alone}, times an uncertain beaming factor that might reduce
the energy by up to 100 at the expense of requiring many more events.

This developing paradigm was challenged last month by the discovery (Galama
et al. 1998ab) of a supernova, SN 1998bw, Type Ib (Sadler et al. 1998; Lidman
et al. 1998) and later Ic (Patat \& Piemonte 1998), within the 8 arc minute
error box of GRB 980425 (Soffita et al. 1998). Extrapolation of the supernova
light curve implied an explosion time consistent with the GRB, an extremely
unlikely occurrence unless the two were associated. Further, the supernova
was unusual, presenting a radio luminosity 100 times brighter than typical
Type Ib's, brighter in fact than any supernova ever observed before (Wieringa
et al. 1998). Moreover, relativistic expansion was inferred (Kulkarni et al
1998), the spectrum was unusual (Lidman et al. 1998; Patat \& Piemonte 1998)
and the light was curve brighter (Galama et al. 1998b) than typical for a Ib
or Ic. {\sl In toto} the case for a GRB-supernova association is compelling.

However, the redshift to the barred spiral galaxy where the supernova
occurred is only 0.0085 (Tinney et al. 1998) and the burst was not an
extraordinarily bright one. The duration and count rate for Beppo Sax were in
fact comparable to GRB 971214 at a red shift of 3.42. From this we infer that
the gamma-ray burst, which lasted 30 seconds, had an energy that was about
10$^{48}$ erg, or 10$^3$ - 10$^4$ times fainter than a typical cosmological
GRB. The BATSE detector on Compton Gamma-Ray Observatory also saw the burst
(Galama et al. 1998b) for about 35 s and inferred a total energy of $8.5 \pm
1.0 \times 10^{47}$ erg in gamma-rays. BATSE saw no emission above 300 keV
for this burst, making it another example of the so called ``no high energy"
GRB's, about 25\% of the BATSE sample. At this luminosity, other GRB's like
GRB 980425 would have been invisible had they occurred 20 times farther away,
so unless this was an extremely serendipitous observation, there must be a
very high spatial density of these events, perhaps thousands of times that of
the ``classical" BATSE bursts ({\sl modulo} the beaming factors). This
requires a source that is very common in nature. Indeed, since BATSE
observations can be explained with an event rate of 10$^{-7}$/yr/L$_*$ galaxy
(Wijers et al. 1998), this suggests an event at least 1\% as frequent as
supernovae.

In order to explain the brilliance of SN 1998bw, if it is powered by
the decay of radioactivity like other Type I's, it is necessary to
synthesize and eject $\gtaprx 0.45$ M\sun \ of $^{56}$Ni in the
explosion. Since this was not a Type Ia supernova - and we know of no
way to make GRB's or strong radio sources based on Type Ia supernovae -
we invoke massive stars. But then the large $^{56}$Ni mass requires,
in traditional models, both a very massive star and a high explosion
energy. The energy must also be large in order to accelerate the mass
- several times that in a typical Type Ib supernova - to the observed
high velocities and to make the light curve peak in only 17 days
(Galama et al. 1998b). Finally, we are prejudiced by the belief that
GRB's require stars so massive that the neutrino powered ``hot bubble"
mechanism for supernova explosion fails (Woosley 1993). This also
leads us to consider stars whose main sequence mass was over 30 M\sun.

As we were writing our paper, the preprint by Galama et al. (1998b)
appeared which references similar conclusions, at least a massive
stellar explosion with large energy, reached in a paper by Iwamoto et
al. (1998). We have not seen that paper and our work has proceeded
independently.

In the following sections we describe the modeling of the supernova
explosion, calculate the fraction and energy of relativistic mass
ejected, and examine the model light curve and spectrum. We also
attempt to understand how the supernova might have made a GRB. The
interaction of the supernova shock with circumstellar material has an
appealing physical basis and might be expected to occur frequently,
but the gamma-ray energy requirements even for this faint burst are
large and are not obtained (in spherical symmetry) even for very
violent explosions. We do find models that agree well with
the multi-band photometry of the supernova and from these are able to
make predictions about the spectrum - unknown to us as of this
writing.

The large explosion energy and lack of a straightforward way of making the
GRB in spherical symmetry suggest that something unusual happened in SN
1998bw. In the Conclusions we discuss what it may have been.

\section{Simulations}

\subsection{The Explosion}

The models we use, which might eventually be tuned to give better
agreement, are based upon massive stars - 25 - 35 M\sun \ on the main
sequence, that have lost their hydrogen envelope and perhaps even
their helium shell.  For 25 M\sun, this may require membership in a
close binary; for 35 M\sun, radiative mass loss will suffice. Once the
helium core is uncovered rapid mass-dependent mass loss may commence
(Langer 1989) that removes a portion of the helium shell. We thus
experiment with both the helium cores and the carbon-oxygen cores of
these massive stars. All calculations of the explosion and expansion
were carried out using the KEPLER code (Weaver, Zimmerman, \& Woosley
1978). The light curve and approximate spectra are calculated using a
different approach ($\S$3.4, 3.5).

Our first model uses the 9.12 M\sun \ helium core of a 25 M\sun \ main
sequence star similar to the one evolved to presupernova by Woosley
\& Weaver (1995). Because we are interested in the correct density
distribution in the atmosphere of the star (for shock acceleration),
it was important that the surface of the helium star be fine zoned and
in thermal and hydrostatic equilibrium. It takes time for the star to
relax into this equilibrium and this cannot be accomplished by a star
that is already exploding. So rather than try to make a "stripped
down" helium core, we used the 25 M\sun \ model at carbon ignition to
construct our model. The hydrogen envelope was removed (down to a
hydrogen mass fraction of 0.01) and the rezoner allowed to prepare a
very finely zoned surface as the outer helium layer expanded. A
surface boundary pressure of 10$^8$ dyne cm$^{-2}$ was necessary to
keep the star numerically stable. This did not appreciably affect the
structure. 10$^{-5}$ M\sun \ (8 zones) into the atmosphere, the
pressure exceeded this boundary value by 10 and the radius had
decreased by only 9\%. This boundary pressure was of course removed
when the star exploded. The outer zone was $2 \times 10^{-6}$
M\sun. This atmosphere was allowed to relax into thermal and
hydrostatic equilibrium and the star was then evolved, without farther
mass loss, through neon, oxygen, and silicon burning to the
presupernova state.  As a presupernova, the star had a luminosity of
$1.8 \times 10^{39}$ erg s$^{-1}$ and radius $2.5 \times 10^{11}$
cm. As before, the iron core mass was 1.78 M\sun. This star was then
exploded using a piston as described in Woosley \& Weaver (1995). The
final kinetic energy at infinity was varied (Table 1). This series of
models is called HE9 with a suffix to indicate explosion energy.

Three other presupernova models were similarly constructed. The next used
the 6.55 M\sun \ carbon-oxygen core of the 25 M\sun \ star at carbon
depletion. Fine surface zoning was again engineered with outer zones
typically $\sim 10^{27}$ g. The radius of the star at explosion was 1.22
$\times 10^{10}$ cm and the luminosity $6.6 \times 10^{38}$ erg s$^{-1}$.
Models from this series are denoted CO6. Two additional models were extracted
from a 35 M\sun \ star at carbon ignition. This gave a helium core of 14.13
M\sun \ (Models HE14) and a carbon-oxygen core of 11.03 M\sun \ (Models CO11).

These models were all exploded using pistons parameterized so as to
give a specified kinetic energy at infinity for the ejecta (Woosley \&
Weaver 1995). Typical values for the ``$\alpha$'' parameter were 10 -
20. The piston was located at the edge of the iron core in each case
(1.78 M\sun \ in the 25 M\sun \ derived models; 2.03 M\sun, for the 35
M\sun \ derived models).  Nucleosynthesis was followed as in Weaver,
Zimmerman, \& Woosley (1978) using the nuclear reaction set described
in Woosley \& Weaver (1995).  The final kinetic energies and
abundances of $^4$He, $^{16}$O, $^{28}$Si, and $^{56}$Ni are given in Table
1.

\section{Observational Properties}

\subsection {Shock Break Out}

The first model ever proposed for gamma-ray bursts was supernova shock
break-out (Colgate 1969, 1974). The outer layers of the star are heated by
the eruption of the strong shock wave, then release their energy as the
layers expand. We followed here the emergence of the shock using the KEPLER
hydrodynamics code (Weaver, Zimmerman, \& Woosley 1978) and a simple
prescription for the opacity - electron scattering based upon a full solution
of the Saha equation (Ensman \& Woosley 1988). As previously noted, the
zoning of the outer layers was very fine, logarithmically smooth down to
10$^{-6}$ M\sun. The radiation transport for this early stage was calculated
using a simple single temperature model of flux-limited radiative diffusion.
It is known that this approach underestimates the temperature of the burst by
up to about two (Ensman \& Burrows 1992), but the luminosity should not be
far off and both should suffice for present purposes.

The results for a representative sample of our models are given in Table 2.
Typical burst luminosities are 10$^{43}$ - 10$^{44}$ erg s$^{-1}$ for up to
several seconds. Typical photon energies (3 kT) are about a keV. While bright
and potentially detectable, this burst of radiation is over two orders of
magnitude softer and four orders of magnitude fainter than what was seen in
GRB 980425.

\subsection {Relativistic Mass Ejection?}

As the shock progresses through the outer layers of the star, it
accelerates.  If the density gradient is steep enough and the shock
strong enough, a portion may even become become relativistic.
Analytic solutions of ultra-relativistic shocks and semi-analytic
solutions of mildly relativistic shocks exist (Johnson \& McKee 1971;
McKee \& Colgate 1973; Gnatyk 1985).  For an exponentially declining
density profile, the product of the Lorentz factor ($\Gamma$) and the
velocity of the shock ($\beta=v/c$ where $c$ is the speed of light) is
given by (Gnatyk 1985):
\begin{equation}
\Gamma \beta \propto (\rho r^{N+1})^{-\alpha},
\end{equation}
where $N$ is a geometric factor set to 2 for spherical symmetry, and
$\alpha$ is determined, via simulations, to be $\sim$0.20.

We can use this scaling relation to estimate the energy ejected as a
function of $\Gamma$ for lower mass zones than we are able to carry in
our present (Newtonian) hydrodynamical calculation. In Figs. 1 and
2 the quantities $\rho r^3$ and Q = $\Gamma \beta (\rho r^3)^{0.2}$ are
plotted as functions of the mass outside of radius $r$. The density
and radius are evaluated in the presupernova star; $\Gamma$ and
$\beta$ are evaluated after the matter has reached the coasting
phase. The scaling relation for $\Gamma$ is not precise because it neglects
the internal energy deposited by the shock and the subsequent acceleration
that energy causes (Fryer \& Woosley 1998a). However the near constancy of Q
suggests that we can extrapolate the well determined sub-relativistic
solution calculated here to higher $\Gamma$'s.

Taking a representative range of Q $\approx 3 - 4 \times 10^5$ and a
scaling relation between $\rho r^3$ and external mass M = 10$^{32}
(\rho r^3/10^{32})^{4/3}$ (Fig. 1), we estimate the kinetic energy,
$\Gamma$Mc$^2$, contained in material having $\Gamma \gtaprx 10$ to
be 10$^{41}$ - 10$^{42}$ erg. For $\Gamma$ of 3 the range is 10$^{44}$
- 10$^{45}$ erg. This is several orders of magnitude less than
required to produce the GRB.

Sub-relativistic matter is also unlikely to produce the burst. To
carry 10$^{48}$ erg requires a minimum of $\sim 10^{27}$ g. This
matter will interact with its own mass before giving up its
energy. For a mass loss rate of 10$^{-5}$ M\sun \ y$^{-1}$ and speed
10$^8$ cm s$^{-1}$, the radius where this will happen is at least
$\sim$10$^{14}$ cm. The light crossing time for this region is
$\gtaprx$10$^4$ s, so the burst would be too long and faint. Raising the mass
loss can give a smaller interaction radius and shorter burst, but at the
expense of becoming optically thick to the gamma-rays that are produced. It
seems likely that an enduring hard x-ray flash will be created - an analogue
to what was seen in SN 1993 J (Leising et al. 1994; Fransson, Lundquist, \&
Chevalier 1996). This lasted about a hundred days at 50 - 100 keV.

An additional concern is that the radio emission implies relativistic
expansion even days after the GRB occurred (Kulkarni \& Frail
1998). There is roughly $5 \times 10^{49}$ erg in the outer 10$^{-3}$
M\sun \ of ejecta of our models here, all moving at about 1/3 c. This
could certainly provide a bright radio source, but the expansion would
not be relativistic.

\subsection{The Supernova Light Curve}

UBVRI photometric observations of SN 1998bw have been reported by Galama et
al. (1998b) and show the supernova falling in brightness when first observed
(0.6 days after the GRB 980425)), then rising to a maximum of $M_V=-19.4$
(a distance of 36 Mpc based on the object's redshift, H$_0 = 70$
km~s$^{-1}$~Mpc$^{-1}$, and $A_V$ = 0.2~mag, is used throughout this
discussion). We have used these observations to estimate the ``bolometric
luminosity" ($L_{UVOIR}$) by integrating over the $UBVRI$ photometry. To do
so, we extend the spectrum beyond the I-band using a blackbody tail and
beyond the U-band with a spline. The results are not sensitive to the
treatment of the infrared, but there is some ambiguity in the treatment of
the ultraviolet. Our procedure here is influenced by previous analyses of
supernovae that had broad wavelength coverage (e.g., Type Ic SN~1994I). Type
I supernovae of all subclasses are affected by line blanketing and it is
important to cut off the ultraviolet spectrum quickly relative to the best
fitting blackbody. The photometric evolution of this object is consistent
with other objects which have a rapidly falling ultraviolet spectrum. The
derived bolometric flux would only be in significantly error if there were a
large amount of flux below 3000 \AA . This appears unlikely except at the
earliest times (less than three days after the GRB). The derived bolometric
light curve is given in Table~3 and in Figs. 3 and 4.

These observational data were used to discriminate among possbile models.
Each model was evolved with the KEPLER hydro-code to 10$^5$ seconds after
explosion at which point a link was made to a multi-group radiation transport
code, EDDINGTON (Eastman \& Pinto 1993). This code solves the time-dependent
transport equation, in the co-moving frame, simultaneously determining the gas
temperature by balancing heating and cooling.  The heating rate includes
energy deposition by gamma-rays from radioactive decay. Gamma-ray transport
was computed using a single energy group approximation to compute the
transport each of gamma-ray line (Woosley et al. 1994).

For the EDDINGTON light curve calculations, the KEPLER grid, which consisted
of 370 to 700 zones, was remapped onto a grid of 80 zones. The
composition was artificially ``moderately mixed", which is to say a running
boxcar average using a grid 1 M\sun \ wide was calculated sliding the grid
out through the star. For those models that had a helium shell, this was not
sufficient mixing to bring $^{56}$Ni up into the helium. Bringing $^{56}$Ni
into the helium layer would probably produce a Type Ib, not Ic supernova
(Woosley \& Eastman 1997).

The opacity included contributions from He I-II, C I-VI, O I-VIII, Si I-X, S
I-X, Ca I-XII, Fe I-XIV, Co I-XIV and Ni I-XIV. Processes included inner
shell and valence shell photoionization, bremsstrahlung, electron scattering,
and line opacity from 90,000 lines, which was represented using the expansion
opacity described by Eastman \& Pinto (1993).

The light curve calculations assumed local thermodynamic equilibrium (LTE).
Gas excitation and ionization was computed by solving the Saha-Boltzmann
equation at the local temperature and density. Because the density is so low
here, the assumption of LTE is questionable. This asuumption remains
approximately valid because the gas is radiatively driven into thermal
equilibrium. But as the ejecta becomes more transparent, the assumption of
LTE gets progressively worse. In general, we find that LTE tends to
overestimate the population of excited states, underestimate the
ionization, and underestimate the gas temperature.

For the present light curve calculations (Figs 3 - 5), the frequency grid
consisted of 500 groups covering the range $30< \lambda < 5\times10^4$
angstroms. Because of this low resolution, spectral features computed by the
light curve code are smeared, but the spectrum is still adequate
for photometry.

The best fit to the light curve and photometry is for our lowest mass,
highest energy explosions (Table 1), those based on the 6 M\sun \
carbon-oxygen core. Even these models do not rise fast enough to agree
with observations during the first few days. More mixing of $^{56}$Ni
to nearly the surface of the explosion would give a more rapid rise,
but in one dimension this mixing would keep a larger volume hot and
ionized at late time and increase the photospheric radius. This would
make the supernova too red. Another possibility is that the
pre-explosive star had a helium layer and a larger radius. The release
of shock deposited energy by helium recombination would then give a
brief ``plateau'' in the light curve as is often calculated for Type
Ib models. There are some indications in the data of the first few
days that the supernova initially faded slightly. This would be
consistent with helium recombination. Alternatively the explosion was
not spherically symmetric ($\S$5).

\subsection{The Supernova Spectrum}

In order to evaluate the effects of the LTE approximation and low frequency
resolution, we carried out a higher resolution, non-LTE calculation of
the spectrum of Model CO6C near maximum light (Fig. 6; 14.4 days). This
calculation assumed steady state between energy deposition and emission.
Gamma-ray transport was computed with the Monte Carlo code FASTGAM (Pinto \&
Woosley, 1988) using a frequency grid of 30,000 groups and a spatial grid
of 41 radial zones. Ions included were He I-II, C I-IV, O I-IV, Si I-IV, S
I-IV, Ca I-IV, Fe I-IV and Co I-IV. The broadband photometry predicted by
this Model was shown in Fig. 5 as solid points.

The agreement with the observations is much improved over the predictions of
the LTE  calculation.  In particular, the predicted U band flux is
a magnitude brighter in the non-LTE calculation. Fig. 6 shows the
spectrum predicted by the non-LTE calculation of Model CO6C just prior to
peak light, (the calculation is at 14.4 days). Although we have not yet had
access to any optical spectroscopy of SN 1998bw, the maximum light spectrum
of CO6C has many of the properties displayed in the maximum light spectrum
described by Patat \& Piemonte (1998): it peaks near 5400 angstroms and shows
strong absorptions by C II, O I, O II, Si II, S II and Ca II. The model does
have a He~I $\lambda5876$ absorption feature, which Patat says was not
present in SN 1998bw. However, it is weak, highly blue shifted, and could
easily be mistaken for something else. Also, the He I $\lambda6678$ is very
weak in Model CO6C, and blended with C II and O II, consistent with Patat's
report on SN~1998bw.

The velocities here are very high. In the unmixed model, most of the
helium (which came from photodisintegration in Model CO6C) was moving
between 0 and 12,000 km s$^{-1}$; carbon was appreciably abundant (over
1\% by mass) only at speeds greater than 25,000 km s$^{-1}$; oxygen was
abundant over 14,000 km s$^{-1}$; magnesium, 15,000 km s$^{-1}$ and up;
silicon 12,000 to 26,000 km s$^{-1}$; calcium, 12,000 - 15,000 km
s$^{-1}$; and cobalt ($^{56}$Ni) was found between 0 and 14,000 km s$^{-1}$.
This inverted speed distribution for helium and heavier elements might be a
distinctive feature in the spectrum of a CO-explosion as opposed to that of a
helium star. In a helium star there might be a bimodal distribution of helium
in velocity. In a CO star high velocity helium is weak (arbitrarily we
defined the outer boundary of the CO model as where helium went to 1\% by
mass in the 25 M\sun \ star igniting carbon burning). The velocities
here are higher than reported by Patat \& Piemonte (1998).

In a later paper, when spectroscopic data is available, we hope to
treat the spectral evolution of SN 1998bw in greater detail. However,
from the information at hand it seems that, photometrically at least,
SN 1998bw is well modeled as the explosion of a carbon-oxygen core of
6 M\sun \ with a kinetic energy of $\sim 2 \times 10^{52}$ erg which
naturally yields a $^{56}$Ni mass near 0.5 M\sun.  The fact that we
used a CO core without an appreciable layer of helium still in place
is in part an expedient. It may well be that a helium core of the same
mass and explosion energy would have worked just as well. If detailed
spectroscopic analysis shows that the high velocities, e.g., of Model
CO6C, are not present, this may indicate an asymmetric explosion
($\S$5).

\section{Other Supernovae}

So why have we not observed events like this before? Or have we? Wang and
Wheeler (1998) have compared the correlation of supernovae with GRB's
and find a positive correlation with Type Ic's, but no correlation between
GRB's and other supernovae.  There have only been 16 supernovae classified as
Type Ic during the six year period 1992-1997 as listed in the Asiago Catalog
(Barbon et al. 1993). Presumably many others have been missed, but they do
not affect the argument. the BATSE sky coverage is about one third, so one
might expect about 5 SN~Ic-GRB correlations if all SN~Ic are GRBs.  But there
may also be considerable variation in the GRB's from supernova to supernova.
Perhaps only the stars with the highest mass and biggest explosion energies
make a visible GRB, or maybe they must be observed from a certain angle.
Nevertheless, it would be interesting to search the known GRB error boxes for
subsequent supernovae - but when would the supernova be discovered?  Two
weeks later, a month?

We checked only three cases because we knew them to be unusually luminous
Type Ic's. These were SN 1992ar (discovered as part of the Calan/Tololo
survey, Hamuy \& Maza 1992); SN 1997cy, (discovered as part of the Mount
Stromlo Abell Cluster supernova search, Germany et al. 1997); and SN 1997ef
(Nakano \& Sano 1997). SN 1992ar was discovered in late July, 1992 and GRB
920616 occurred about two sigma from the SN's position. SN 1997ef, discovered
on November 25, 1998, has also been pointed out by Wang and Wheeler along
with its coincidences (within 3-sigma error boxes) with GRB 971115 and GRB
971120. While it is interesting that both of these supernovae have a
reasonable GRB candidate, neither is a particularly compelling case because
of the large separation between GRB and the centrod of the error box.
However, the situation is different for SN 1997cy. This supernova (not in
Wang and Wheeler's list) had a bizarre spectrum, with broad Ic-like lines
like observed in SN 1997ef and 1992ar, but also a H$\alpha$ line with broad
and narrow components. SN 1997cy was also the most luminous supernova ever
discovered, having $M_R\approx -21$ at maximum. GRB 970514, a burst with a
smaller than typical error box (3-degrees), occurred less than a degree away
at a time compatable with the discovery and pre-discovery images. This object
is the subject of a paper by Germany et al. (1998). So perhaps SN 1998bw is
not an isolated case.

However, we want to state clearly that we do not believe that all or even a
majority of Type Ib (or Ic) supernovae make GRB's. Most of these supernovae
are very well modelled by a lower mass explosion (3 - 4 M\sun \ helium core)
that makes about one-third as much $^{56}$Ni as SN 1998bw and expands with
moderate energy $\sim$10$^{51}$ erg (e.g., Woosley \& Eastman 1997). Even the
more massive stars and unusual explosions studied here might only make a GRB
when viewed at certain angles. As we discuss in the next section, the GRB is
probably beamed while the supernova is certainly visible at all angles. We
expect a GRB supernova association only in the unusual case and two-thirds of
these will be missed by BATSE.

\section{Conclusions}

SN 1998bw was and continues to be an unusual supernova. When modeled
as a spherically symmetric explosion, it requires an energy over 20
$\times 10^{51}$ erg, a $^{56}$Ni mass over 0.45 M\sun, rapid
expansion, high stellar mass, and high mass loss rate (to explain the
radio). Of course the most unusual property of SN 1998bw was its
proximity to GRB 980425. We have assumed here that the two are related
and have looked for ways the supernova might make the burst. For our
one-dimensional models we found none.

However, we do find good agreement with the multiband photometry of
Galama et al. (1998b), and the bolometric light curve integrated from
that data, and the explosion of a 6 M\sun \ core of carbon, oxygen,
and heavy elements with final kinetic energy $2 - 2.5 \times 10^{52}$
erg. The explosion leaves behind a 1.78 M\sun \ (baryonic mass)
object, presumably a neutron star and makes about 0.5 M\sun \ of
$^{56}$Ni. However the mass of the remnant and the explosion energy
were not calculated in a consistent way, but were free parameters. We
do not think it is critical that our best fit was a carbon-oxygen core
and not a helium core; the key quantity is the energy to mass
ratio. Type Ic supernovae have weak helium lines chiefly as a
consequence of weaker mixing between the helium and $^{56}$Ni shells
than in Type Ib (Woosley \& Eastman 1997). Even this very energetic
explosion is too faint the first few days of the supernova. There are
two explanations for this - either there was a helium layer with a
larger photospheric radius than the carbon-oxygen core used here that
gave a brighter ``plateau" before the radioactive decay energy
diffused out, or the explosion was asymmetric, ejecting some $^{56}$Ni
almost to the surface at some angles - a very mixed model. However,
spherically symmetric mixing would have given a larger photosphere and
perhaps a redder supernova than was observed (Woosley \& Eastman
1997). Helium may be present in the spectrum even in our carbon-oxygen
core models, but it is chiefly from photodisintegration and would be
the slowest not the fastest moving ejecta. High velocity helium would
be a signature of a helium star.

All in all, though the parameters may be extreme, especially the
explosion energy, one could model SN 1998bw in a qualitatively similar
way to other Type Ib and Ic supernovae, that is if it were not the
origin of GRB 980425.

But we believe that it was. So what happened? Can nature really
provide $2 \times 10^{52}$ erg to a supernova whose main sequence mass
was over 25 M\sun? Current belief (e.g., Burrows 1998; Fryer 1998) is
to the contrary. If anything, the explosion actually becomes weaker as
one goes to larger mass. The iron core is larger and can potentially
provide more neutrinos, but it is also close to criticality and the
mass flux from the imploding mantle of the star is formidable. It is
very difficult to stop the implosion before the neutron star gives way
and collapses to a black hole.

And so it may be that something else happened here, that the explosion
was not spherical and powered by neutron star formation, but very
asymmetric and powered by jets from black hole formation. Bodenheimer
and Woosley (1983) first considered such an outcome to black
hole formation and found that a supernova still resulted. Woosley
(1993) and Hartmann \& Woosley (1995) emphasized jet production and
proposed an association of this model with gamma-ray bursts.  Initially this
model was referred to as the ``failed supernova'' (because
the prompt supernova mechansim failed), and later as the ``collapsar model''
(Woosley 1996), because it was the outcome of a collapsed star. A model
having very similar characteristics, called the ``hypernova'', has been
discussed by Paczynski (1997). Fryer \& Woosley (1998b) have also
discussed setting up very similar conditions in the merger, by common
envelope, of a stellar mass black hole and the helium core of a
massive supergiant star. Current two dimensional studies of the collapsar
model by MacFadyen and Woosley (1998ab) are encouraging. Specifically they
find, in the collapse of a 14 M\sun \ rotating helium star to a black hole,
an accretion rate of over 0.1 M\sun \ s maintained for about 10 s as the
black hole grows from 2 M\sun \ to 7 M\sun. The Kerr parameter, $a$, grows to
$\gtaprx$ 0.9 early on. For these conditions, Popham, Woosley, \& Fryer
(1998) find that the annihilation of neutrinos radiated from the viscous disk
deposits up to 10$^{51}$ erg s$^{-1}$ along the rotational axis of the black
hole. Thus energies as much as 10$^{52}$ erg are potentially available.  Some
of this energy goes into accelerating relativistic jets along the rotational
axes, but more may go into ejecting a lot of mass at lower speeds. MacFadyen
\& Woosley have not calculated the evolution beyond 15 s.  Large amounts of
energy can also potentially be extracted from the rotation of the black hole
(e.g., Meszaros \& Rees 1997).

Viewed this way, GRB 980425 was a low energy analogue of the
enormously more luminous ``classic" GRB's. Both are produced by black
hole accretion, but in GRB 980425 the jet energy was weaker and
$\Gamma$, along our line of sight, lower.  Perhaps if we had viewed GRB
980425 straight down the axis a more powerful, harder GRB would have been
seen. Or maybe the helium core mass, rotation rate, and therefore black hole
accretion rate were not so extreme in GRB 980425 as in other GRB's. Viewed
from the side though, in any case, the emission from the high $\Gamma$ jet
would have been supressed and spread over a much longer time, probably
invisible. But even at our angle there may have been, say, 10$^{-7}$ -
10$^{-6}$ M\sun \ moving with $\Gamma \approx$ 10. Colliding with the
pre-explosive mass loss at about 10$^{13}$ - 10$^{14}$ cm, this would have
made the observed burst (Meszaros \& Rees 1993). If we had seen SN 1998bw at
still lower latitudes, the GRB would have been missed.

Once spherical symmetry is abandoned an entirely different solution
becomes possible for the supernova. If matter can fall in to close to
the black hole and come out again (MacFadyen \& Woosley 1998b), the
production of $^{56}$Ni is not directly tied to the shock energy and
pre-explosive density structure of the star. It is as if $^{56}$Ni
could be made ``convectively''. The one number we view with some
confidence here is that SN 1998bw made about 0.5 M\sun \ of
$^{56}$Ni. But suppose it could do so while only ejecting a few
solar masses of heavy elements and helium. Then the correlation
between $^{56}$Ni mass and explosion energy is lost. SN 1998bw could
have been a slower moving, lower energy explosion (shared by a smaller
ejected mass) than we have calculated here and still have peaked as
early as it did.

It is unfortunate that so many questions remain unresolved. First, is
it certain that SN 1998bw and GRB 980425 are the same thing? Future
missions with smaller error boxes (e.g., HETE- 2) should show if this
is the case. Finding other historic Type Ic supernovae in coincident
with GRB locations from BATSE would also lend credence to this
identification. We have given two possible examples. There may be
more.

Can a combination of theory and obsevation still tell us what happened
in this supernova/GRB? Continued spectroscopic monitoring of the
supernova will obviously be an important diagnostic as the supernova
enters (has in fact already entered) its nebular phase. What widths
and asymmetries are apparent in the lines of oxygen, iron, helium,
silicon, calcium, and carbon?  Are the high velocities of Model CO6C
really there?  What is the mass of the ejecta? Multi-dimensional
modeling of the explosion and radiation transport in the collapsar
model should also show whether it can explain the observations. If it
does not, perhaps something even more interesting has occurred.

\acknowledgements

The authors gratefully acknowldge helpful conversations on the subjects
of gamma-ray bursts and SN 1998bw with Roger Chevalier, Dale Frail, Chris
Fryer, Lisa Germany, Chryssa Kouveliotou, Andrew MacFadyen, Bob Popham, and
Eileen Sadler. This research has been supported by the DOE (W-7405-ENG-48),
NASA (MIT SC A292701), and the NSF (AST-97-31569)

\begin{deluxetable}{lcccccc}
\tablewidth{35pc}
\tablecaption{Explosions Simulated}
\tablehead{\colhead{Model} & \colhead{Mass} & \colhead{Kinetic Energy} &
\colhead{Mass $^{4}$He} & \colhead{Mass $^{16}$O} & \colhead{Mass $^{28}$Si}
& \colhead{Mass $^{56}$Ni} \\
\colhead{} & \colhead{(M\sun)} & \colhead{(10$^{51}$ erg)} &
\colhead{(M\sun)} & \colhead{(M\sun)} & \colhead{(M\sun)} &
\colhead{(M\sun)}}
\startdata
CO6A$^a$ & 6.55   & 5.5 & 0.06 & 3.3 & 0.28 & 0.32  \nl     %co6a1g
CO6B & 6.55   & 15  & 0.14 & 3.1 & 0.36 & 0.42  \nl     %co6a1j
CO6C & 6.55   & 22  & 0.20 & 2.9 & 0.40 & 0.47  \nl     %co6a1l
CO6D & 6.55   & 28  & 0.26 & 2.8 & 0.42 & 0.49  \nl     %co6a1k
CO11A & 11.0  & 9.1 & 0.09 & 6.3 & 0.54 & 0.68  \nl     %co11a1g
CO11B & 11.0  & 25  & 0.21 & 5.9 & 0.70 & 0.84  \nl     %co11a1j
HE9A &  9.12  & 3.7 & 2.4  & 3.0 & 0.35 & 0.51  \nl     %he9a3i
HE9B &  9.12  & 7.7 & 2.4  & 2.9 & 0.39 & 0.58  \nl     %he9a3g
HE9C &  9.12  & 21  & 2.5  & 2.5 & 0.54 & 0.77  \nl     %he9a3j
HE14A & 14.1  & 4.2 & 2.8  & 6.2  & 0.46 & 0.73 \nl     %s35s7b2f
HE14B & 14.1  & 10  & 2.8  & 6.0  & 0.51 & 0.86 \nl     %s35s7b2g

\tablenotetext{a}{``CO" models are carbon-oxygen cores devoid of any helium
surface layer. ``HE" models retain their helium shells.}

\enddata
\end{deluxetable}

\begin{deluxetable}{lccc}
\tablewidth{35pc}
\tablecaption{Shock Break Out}
\tablehead{ \colhead{Model} & \colhead{L$_{\rm peak}$} & \colhead{T$_{\rm
peak}$} & \colhead{Duration}}
\tablehead{ \colhead{} & \colhead{(10$^{42}$ erg s$^{-1}$)} &
\colhead{(10$^6$ K)} & \colhead{(FWHM sec)}}

\startdata
CO6A & 3.0 & 2.2 &  0.24 \nl     %co6a1g
CO6B & 9.1 & 3.0 &  0.11 \nl     %co6a1j
CO6D & 19  & 3.6 &  0.08 \nl     %co6a1k
CO11A & 5.6 & 1.3 & 5.8  \nl     %co11a1g
HE9B &  130 & 1.2 & 4.0  \nl     %he9a3g
HE9C &  270 & 1.4 & 2.5  \nl     %he9a3j

\enddata
\end{deluxetable}

\newpage
\begin{deluxetable}{cc}
\tablecaption{The Bolometric Light Curve of SN 1998bw}
\tablehead{\colhead{Days After GRB Event}&\colhead{$\log(L_{UVOIR})$}\\ & ergs~s$^{-1}$ \\}
\startdata
	   2  &    42.38 \nl
	   5  &    42.63\nl
	  10  &    42.94\nl
	  15  &    43.05\nl
	  20  &    43.00\nl
	  25  &    42.88\nl
	  30  &    42.75\nl
	  35  &    42.63\nl
	  40  &    42.53\\
\enddata
\end{deluxetable}

\begin{figure}
\plotfiddle{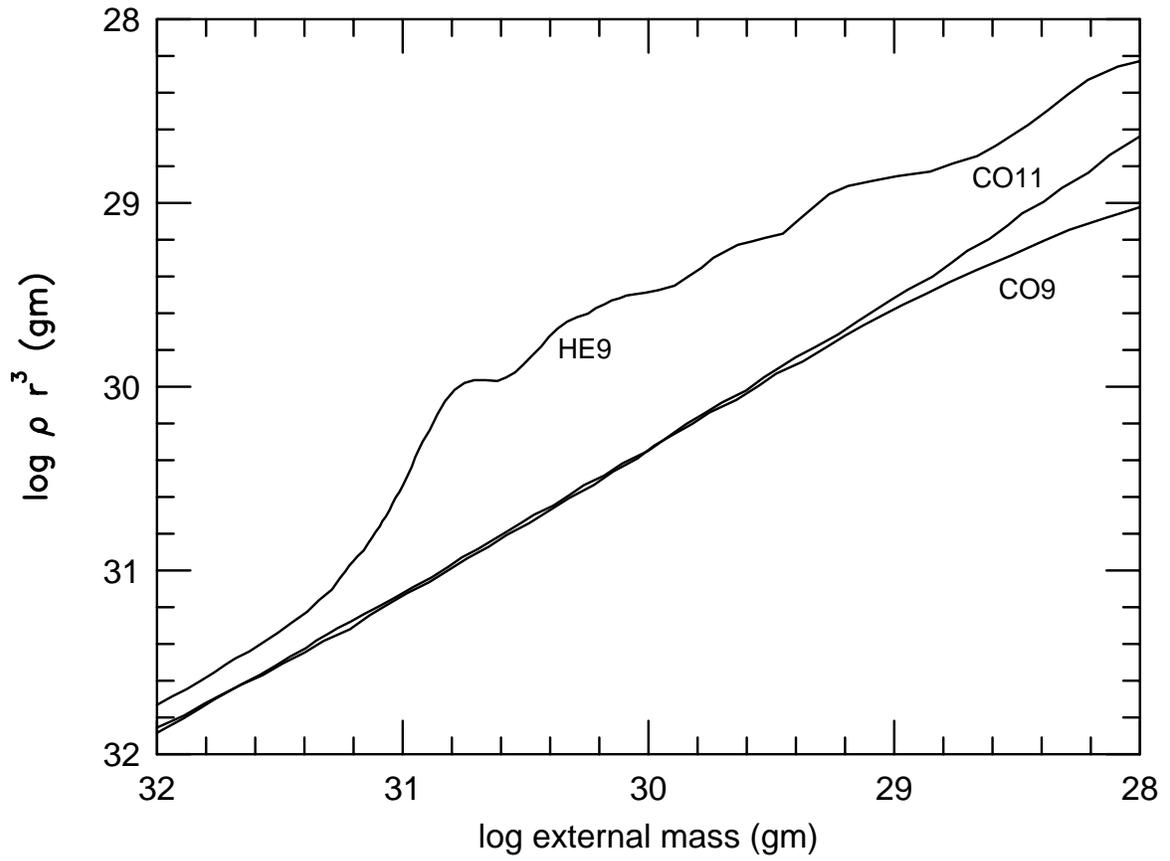}{6in}{0}{70}{70}{-280}{0}
\caption{The quantity $\rho r^3$ is plotted as a function of external mass
for the three pre-explosive models employed in this study. An empirical
relation ${\rm M_{ext}} \propto (\rho r^3)^{4/3}$ is apparent.}
\end{figure}

\begin{figure}
\plotfiddle{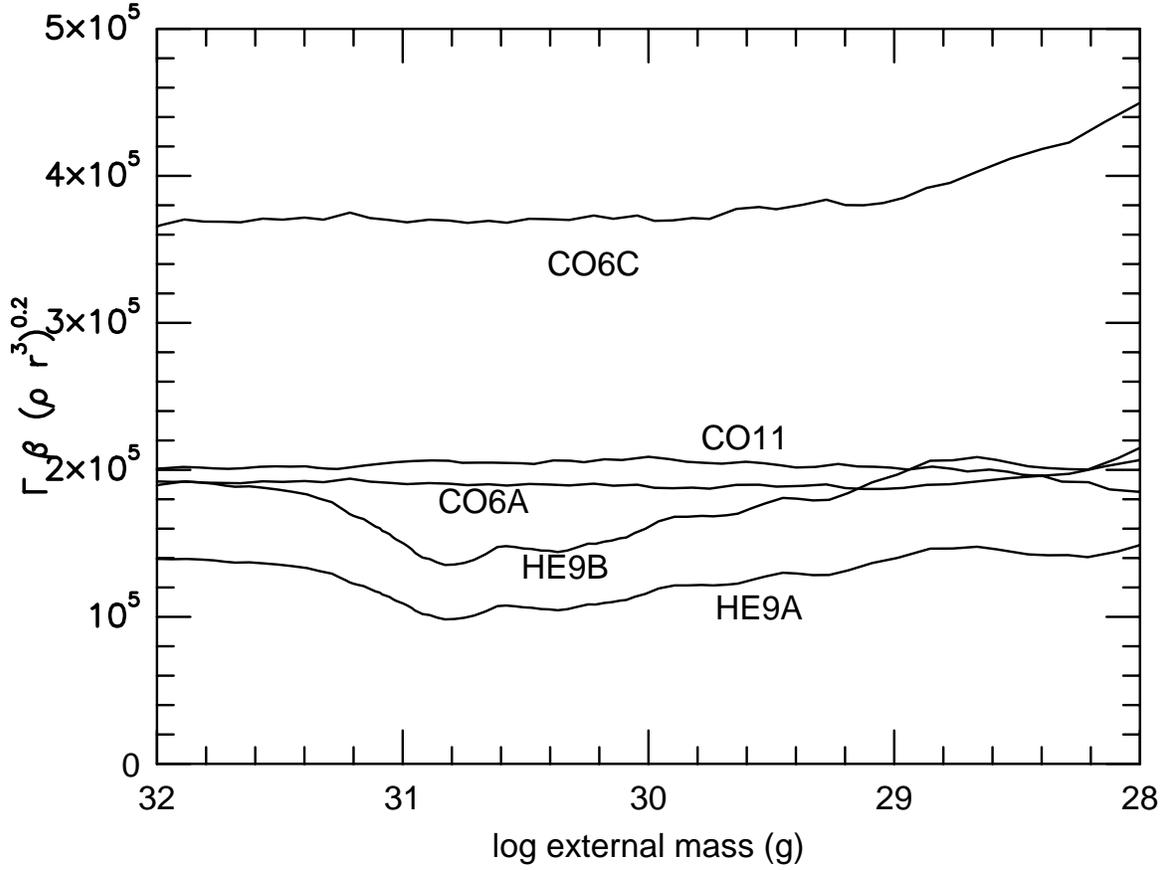}{6in}{0}{70}{70}{-280}{0}
\caption{The quantity $\Gamma \beta(\rho r^3)^{0.2}$ is plotted as a function
of external mass for several runs after they have reached homologous
expansion. Note the near constancy of this product over a large
range in external mass, pre-explosive stellar radius, and explosion
energy. The upturn of some of the models for low external mass is
artificial and a consequence of the velocity approaching the speed
of light in the non-relativistic hydro-code. Scaling this quantity to lower
values of $\rho r^3$ allows us to estimate the energy and mass ejected as a
function of $\Gamma$}
\end{figure}

\begin{figure}
\plotfiddle{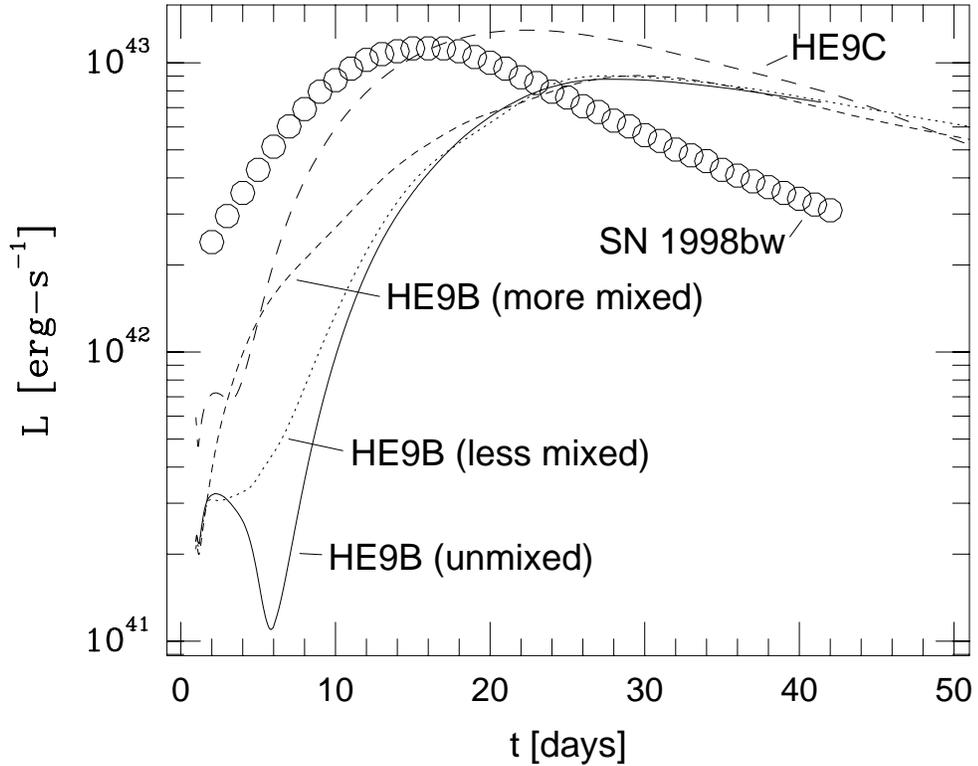}{6in}{-90}{70}{70}{-180}{450}
\caption{The bolometric light curve for the 9 M\sun \ helium
core explosions (Table 1) as calculated using EDDINGTON  compared to the
bolometric light curve obtained by digitizing and integrating the data of
Galama et al. (1998b). The distance is assumed to be 36 Mpc (H$_o$ = 70 km
s$^{-1}$ Mpc$^{-1}$) and the reddening $A_V$ = 0.20. The bolometric data
points are obtained by extrapolating a Planck tail into the infrared and a
spline into the ultraviolet. Even the most energetic HE9 explosions rise too
slowly and peak too late to agree with observations.}
\end{figure}

\begin{figure}
\plotfiddle{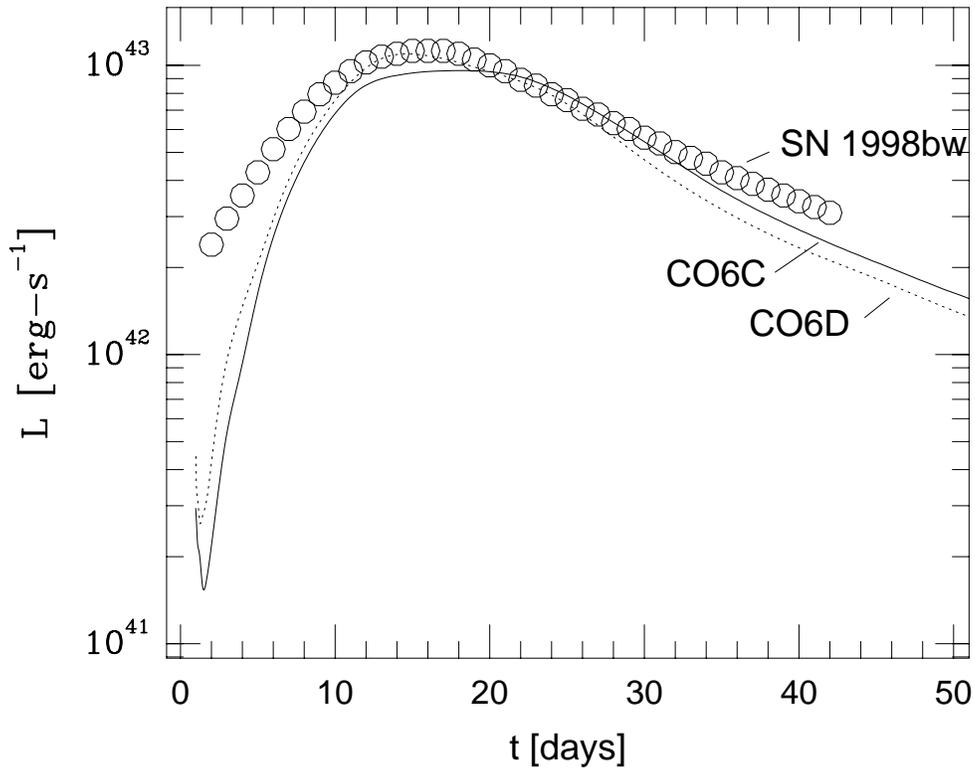}{6in}{-90}{70}{70}{-280}{520}
\caption{The bolometric light curve for the 6 M\sun \ carbon-oxygen
core explosions (Table 1) as calculated using EDDINGTON compared to the
bolometric light curve (see Fig. 3). For Models CO6C and CO6D the agreement
is acceptable, although the models still rise too slowly to explain the
brightness of the supernova during the first few days.}
\end{figure}

\begin{figure}
\plotfiddle{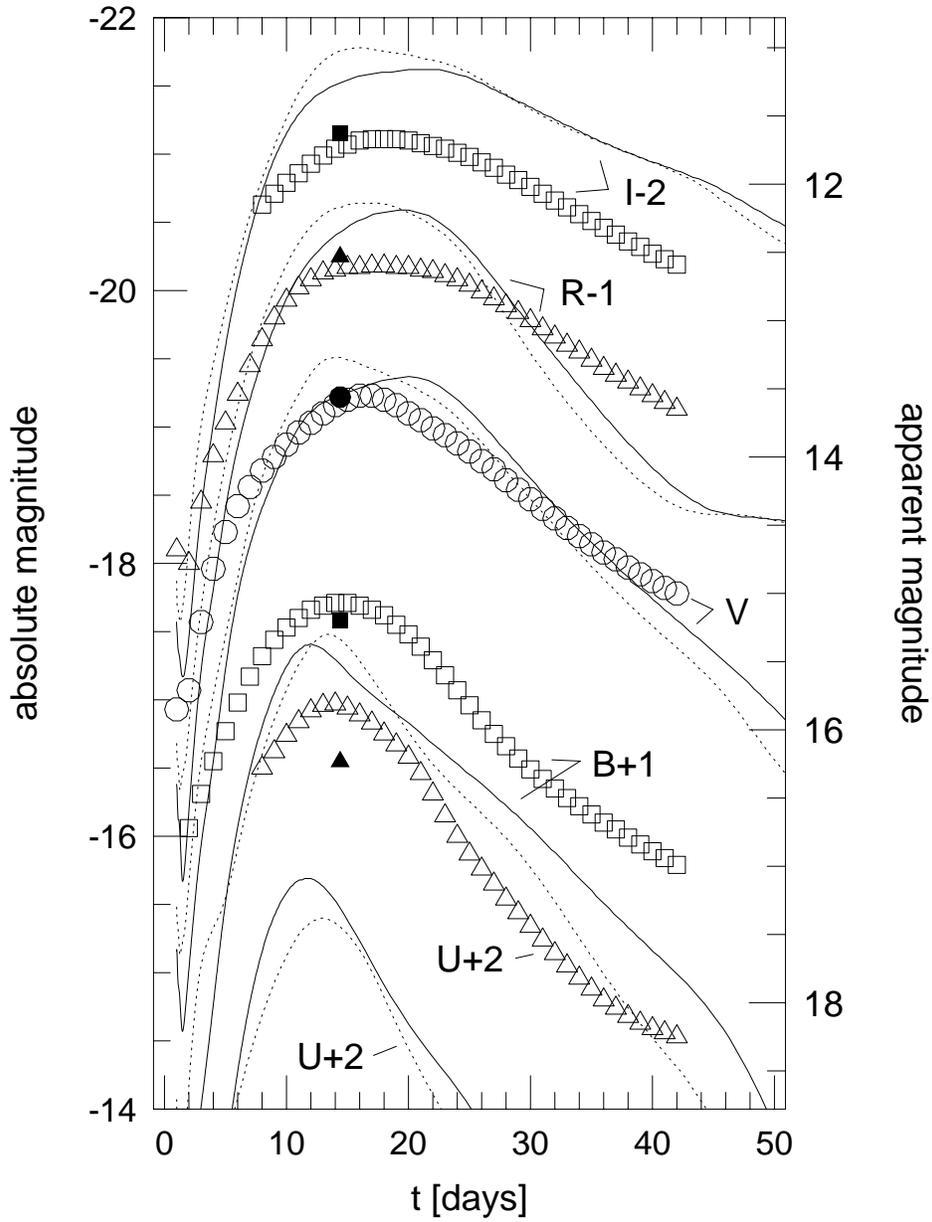}{6in}{-90}{70}{70}{-280}{520}
\caption{The multi-band photometry for Model CO6C as calculated using
the EDDINGTON code compared to the observations of Galama et
al. (1998b). Also given as solid points at maximum light are the
results of a non-LTE spectral calculation of the same model. At least
at peak light, the agreement between the non-LTE calculation and
observations is excellent.}
\end{figure}

\begin{figure}
\plotfiddle{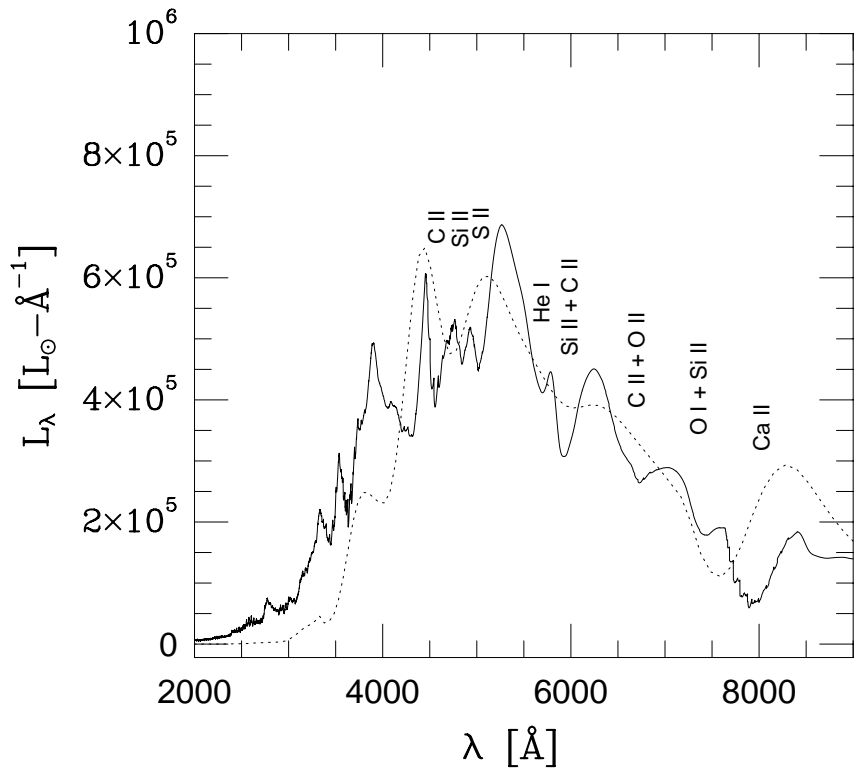}{6in}{-90}{70}{70}{-280}{420}
\caption{The non-LTE spectrum of Model CO6C at maximum optical light 
(solid curve) compared to the LTE spectrum (dashed curve) used to evaluate the
photometric evolution. Both spectra are theoretical. An observed
spectrum was not available at this writing.}

\end{figure}

\end{document}